\pdfoutput=1

\documentclass[aip,jcp,reprint,groupedaddress,footinbib]{revtex4-1}
\usepackage{graphicx} %
\usepackage{amsmath}
\usepackage{mathtools} %
\usepackage{bm}
\usepackage{bbold}

\usepackage{hyperref}
\hypersetup{colorlinks=true,linkcolor=blue,citecolor=blue,urlcolor=blue}

\usepackage{braket} %
\newcommand{\ketbra}[2]{\ket{#1}\!\bra{#2}}

\newcommand{\eu}{\mathrm{e}^}
\newcommand{\iu}{\ensuremath{\mathrm{i}}}
\newcommand{\rmd}{\mathrm{d}}
\newcommand{\half}{{\ensuremath{\tfrac{1}{2}}}}

\newcommand{\op}[1]{\ensuremath{\hat{#1}}}
\providecommand{\mat}[1]{\mathsf{#1}}
\renewcommand{\mathbf}[1]{\bm{#1}}

\DeclareMathOperator{\Tr}{Tr}

\DeclareMathOperator{\Real}{Re}
\DeclareMathOperator{\Imag}{Im}

\DeclareMathOperator{\Ai}{Ai}
\renewcommand{\Re}{\Real}
\renewcommand{\Im}{\Imag}
\newcommand{\T}{{\mathstrut\top}}
\newcommand{\ImF}{\mbox{$\Imag F$}}

\newcommand{\pder}[3][]{\frac{\partial^{#1}{#2}}{\partial{#3}^{#1}}}
\newcommand{\pders}[3]{\frac{\partial^2{#1}}{\partial{#2}\partial{#3}}}

\newcommand{\eqn}[1]{Eq.\,(\ref{#1})}
\newcommand{\Eqn}[1]{Equation~(\ref{#1})}
\newcommand{\eqs}[1]{Eqs.\,(\ref{#1})}
\newcommand{\fig}[1]{Fig.~\ref{fig:#1}}
\newcommand{\secref}[1]{Sec.~\ref{sec:#1}\@}
\newcommand{\Ref}[1]{Ref.~\onlinecite{#1}}

\begin{document}

\title{Semiclassical Green's functions
and an instanton formulation of
electron-transfer rates in the nonadiabatic limit}

\author{Jeremy O. Richardson}
\email{jeremy.richardson@fau.de}
\author{Rainer Bauer}
\author{Michael Thoss}
\affiliation{
Institut f{\"u}r Theoretische Physik und Interdisziplin{\"a}res Zentrum f{\"u}r Molekulare Materialien,
Friedrich-Alexander-Universit{\"a}t Erlangen-N{\"u}rnberg (FAU),
Staudtstra{\ss}e 7/B2,
91058 Erlangen, Germany
}

\date{\today}

\begin{abstract}
We present semiclassical approximations to Green's functions of multidimensional systems,
extending Gutzwiller's work to the classically forbidden region.
Based on steepest-descent integrals over these functions,
we derive an instanton method for computing the rate of nonadiabatic
reactions, such as electron transfer, %
in the weak-coupling limit,
where Fermi's golden-rule can be employed.
This generalizes Marcus theory to systems for which the environment free-energy curves are not harmonic
and where nuclear tunnelling plays a role.
The derivation avoids using the \ImF\ method
or short-time approximations to real-time correlation functions.
A clear physical interpretation of the nuclear tunnelling processes
involved in an electron-transfer reaction
is thus provided.
In the following paper, \cite{GoldenRPI}
we discuss numerical evaluation of the formulae.
\end{abstract}

\maketitle

\section{Introduction}

Electron transfer is a key step in many important molecular processes, %
including redox reactions in electrochemistry
and charge separation in photosynthesis and solar cells. \cite{Marcus1993review,UlstrupBook,ChandlerET}
The electron resides initially on a donor molecule and is transferred to an acceptor,
accompanied by a reorganization of the polar environment.
This reaction can be characterized as a transition between
the donor and acceptor electronic states
with potential-energy surfaces
describing the reactant and product environments.
As such, electron transfer is the simplest example of a nonadiabatic reaction
involving transitions between different electronic states,
requiring a theoretical treatment beyond the Born-Oppenheimer approximation.
\cite{ConicalIntersections1,Tully2012perspective,Kapral2015QCL}
We are thus interested in studying a multidimensional curve-crossing problem,
which as it involves discrete states, is inherently quantum mechanical.
Although we pay particular attention to the electron-transfer problem,
the nonadiabatic formalism is also relevant in other areas of science,
such as defect tunnelling in solids. \cite{EsquinaziBook}
Most formulae derived in this paper can be transferred directly to these fields.

In this paper, we consider the golden-rule limit
which occurs when the transfer of the electron,
rather than the rearrangement of the environment,
is the bottleneck to the reaction. \cite{ChandlerET}
This is quite commonly the case in problems of interest,
especially if the donor and acceptor are spaced far apart.
The standard approach for treating electron-transfer problems in this limit is Marcus theory, \cite{Marcus1985review}
which is based on Fermi's golden rule with the additional approximation that the nuclei are treated classically.
The free-energy curves of the environment in the reactant and product states are also assumed to be harmonic
with the same frequency. %
Although it is quite common for the environment fluctuations to be harmonic, \cite{Kuharski1988Fe3e,King1990ET}
there is no reason for the frequencies to be exactly equal in the two cases,
unless the reaction is symmetric such that the products are equivalent to the reactants.
In some cases the reorganization energies for the forward and backward reaction can differ by a factor of two, \cite{Krapf2012Marcus}
and it is necessary to use a more general rate expression
which allows for this asymmetry but retains the classical and harmonic approximations. \cite{Tang1994asym,CasadoPascual2003ET}

Marcus theory also neglects nuclear quantum effects, such as tunnelling and delocalization,
which have been found to be significant in electron-transfer problems
even at room temperature. \cite{Bader1990golden}
It is however possible to compute the quantum golden-rule rate
exactly for the spin-boson model, \cite{Garg1985spinboson,*Leggett1987spinboson,Weiss}
which treats all environment modes as linearly-coupled harmonic oscillators.
Extensions of this to treat non-linear couplings is also possible. \cite{Tang1994asym}
The multilayer multiconfigurational time-dependent Hartree (MCTDH) method \cite{Wang2003MLMCTDH,*Thoss2006MLMCTDH}
is in principle able to compute the exact rate for such systems, \cite{Wang2006flux}
but in practical applications to large systems is usually limited to
specific forms of the Hamiltonian such as system-bath models.

Thus for many problems in the golden-rule limit,
an accurate calculation of the reaction rate
will require an efficient method
which includes nuclear quantum effects
and treats the potential-energy surfaces in a general way
without making global harmonic approximations.
In this article, we present a new derivation of such a method for computing the rate approximately
using a time-independent picture for the nuclear degrees of freedom.

Our derivation is based on an exact expression for the golden-rule rate
in terms of the Green's functions describing the nuclear quantum dynamics of the reactant and product systems at a given energy.
By extending Gutzwiller's work, \cite{Gutzwiller1967semiclassical,Gutzwiller1971orbits,GutzwillerBook}
we present the semiclassical limit of these Green's functions in the classically forbidden region,
which may also be useful for other applications.
These functions are defined in terms of imaginary-time classical trajectories
which join together to form a periodic orbit, also known as an instanton.
The definition of the rate is found by performing a number of steepest-descent integrations.

Other instanton approaches are well known
from adiabatic rate \cite{Miller1975semiclassical,Coleman1977ImF,*Callan1977ImF,
									Uses_of_Instantons,Affleck1981ImF,Benderskii,rpinst,Althorpe2011ImF}
and tunnelling splitting \cite{Milnikov2001,tunnel}
calculations,
where in both cases the Born-Oppenheimer approximation is first applied to obtain a single-surface Hamiltonian.
The \ImF\ method \cite{Langer1969ImF}
can be used to derive the instanton approximation
to the adiabatic escape rate from metastable states
in the high- or low-temperature limit \cite{Affleck1981ImF}
although its application to finite temperatures \cite{Althorpe2011ImF}
or nonadiabatic reactions \cite{Cao1997nonadiabatic} is less well understood. 
In this approach, the rate
is expressed approximately as the imaginary part of a free-energy,
which is obtained by analytic continuation of a divergent integral. \cite{Kleinert}

Although instanton theory dates back many years,
a numerical method for its efficient application to complex multidimensional systems
has only been developed more recently, \cite{Andersson2009Hmethane,rpinst}
and it is thus experiencing a revival of interest. \cite{Andersson2011HCO,DMuH,Goumans2010Hbenzene,*Goumans2011Hmethanol,*Meisner2011isotope,*Rommel2011locating,*Rommel2011grids,*Rommel2012enzyme,water,*octamer}
The present work can be considered to belong to the same class of methods,
which treat the collective motion of the nuclear modes with a multidimensional instanton.
We note however that our approach is very different from
the semiclassical instanton approach presented in \Ref{Shushkov2013instanton}.
This was derived to recover Marcus theory in both the normal and inverted regimes
using an instanton to describe the dynamics of the transferred electron,
while treating the environment classically.

It has become common to define thermal reaction rates
using time correlation functions. \cite{Miller1983rate}
However, our derivation in energy-space gives us access
not only to the thermal rate but also the energy-dependent reaction probability.
This can be combined with any energy distribution
(including microcanonical)
to give a nonequilibrium instanton rate,
which may be helpful for understanding scattering calculations of large molecules
or gas-phase unimolecular reactions.
The modified correlation function presented in \Ref{nonoscillatory},
which is used for computing nonadiabatic rates avoiding oscillatory functions,
also requires that the flux operator is biased towards energy-conserving electron-transfers.
If we are to combine this method with an instanton approach,
it will be natural to use a formulation in energy-space rather than in time
to access the necessary variables.

An outline of the paper is as follows.
We express the formula for the golden-rule rate in terms of Green's functions in \secref{golden}
and give a semiclassical approximation to these functions in \secref{greens}\@.
Using this approximation, the golden-rule instanton rate is derived in \secref{instanton} in two forms
which are evaluated analytically for a system with linear potentials, the spin-boson model, and in the classical limit for general potentials in \secref{analytic}\@.
\secref{conclusions} concludes the article.
In Paper II, \cite{GoldenRPI} which follows this article,
we discuss how the instanton formula can be applied numerically to complex systems
and relate it to Wolynes' quantum instanton approach. \cite{Wolynes1987nonadiabatic}

\section{Quantum Golden-Rule Rate}
\label{sec:golden}

We consider a general multidimensional system with two electronic states,
each with a nuclear Hamiltonian of the form
\begin{align}
	\label{Hn}
	\op{H}_n &= \sum_{j=1}^f \frac{\op{p}_j^2}{2m} + V_n(\op{\mat{x}}),%
\end{align}
where
$n\in\{0,1\}$ is the electronic-state index
and
$\mat{x}=(x_1,\dots,x_f)$
are the Cartesian coordinates %
of $f$ nuclear degrees of freedom.
These nuclei move on the potential-energy surface $V_n(\mat{x})$
with conjugate momenta $\op{p}_j=-\iu\hbar\pder{}{\op{x}_j}$.
Without loss of generality, the nuclear degrees of freedom have been mass-weighted such that each has the same mass, $m$.
The electronic states $\ket{n}$ are coupled by $\Delta(\mat{x})$ to give the total Hamiltonian
in the diabatic representation, \cite{ConicalIntersections1}
\begin{align}
	\op{H} &= \op{H}_0 \ketbra{0}{0} + \op{H}_1 \ketbra{1}{1} + \Delta(\op{\mat{x}}) \big( \ketbra{0}{1} + \ketbra{1}{0} \big).
\end{align}

The thermal rate, $k$, of transfer of reactants, defined by the projection operator $\ketbra{0}{0}$,
to products, defined by $\ketbra{1}{1}$,
is given exactly by
\begin{align} \label{kthermal}
	k Z_0 = \frac{1}{2\pi\hbar} \int P(E) \, \eu{-\beta E} \, \rmd E,
\end{align}
where $\beta=1/k_\mathrm{B}T$ is the reciprocal temperature %
and $Z_0$ the reactant partition function $\Tr\big[\eu{-\beta\op{H}}\ketbra{0}{0}\big]$.
$P(E)$ is the reaction probability with energy $E$,
defined as \cite{Miller1983rate}
\begin{align}
	P(E) = \half (2\pi\hbar)^2 \Tr\left[ \op{F} \delta(E-\op{H}) \op{F} \delta(E-\op{H}) \right],
\end{align}
where the flux from reactants to products is
\begin{align}
	\op{F}
	&=\frac{\iu}{\hbar} \Delta(\op{\mat{x}}) \big( \ketbra{0}{1} - \ketbra{1}{0} \big).
\end{align}
Note that, although we shall not make use of it in this paper,
it would be simple to replace the canonical ensemble in \eqn{kthermal} with any distribution of energy
to obtain microcanonical or non-thermal reaction rates.

In this paper, we consider only systems for which the electronic coupling, $\Delta(\mat{x})$, is very weak
such that second-order perturbation theory, known as the golden-rule approach, can be employed.
In this limit the formulae reduce to
\begin{align} \label{Pgolden}
	P(E) &= 4 \pi^2 \Tr\left[ \Delta(\op{\mat{x}}) \delta(E-\op{H}_0) \Delta(\op{\mat{x}}) \delta(E-\op{H}_1) \right] \\
	Z_0 &= \Tr\left[ \eu{-\beta \op{H}_0} \right].
\end{align}
These equations, combined with \eqn{kthermal},
give the golden-rule rate in the form also derived from a limit of the non-oscillatory flux correlation function, \cite{nonoscillatory}
and are exact to order $\Delta^2$.

Expanding the trace in \eqn{Pgolden} in the basis of reactant and product internal (e.g.\ vibrational) states $\ket{\mu}$ and $\ket{\nu}$
would give the standard golden-rule formula, \cite{Zwanzig}
\begin{align}
	k Z_0 = \frac{2\pi}{\hbar} \sum_{\mu\nu} \left|\braket{\nu|\Delta(\op{\mat{x}})|\mu}\right|^2 \delta(E_\mu-E_\nu) \, \eu{-\beta E_\mu},
\end{align}
where the sum over states should be replaced by the integral $\iint \rmd E_\mu \rmd E_\nu$ for continuous systems.
Considering this golden-rule formula,
we notice that the reaction occurs between internal states of equal energy.
For the majority of systems of interest (in the Marcus normal regime),
at low enough temperatures %
the dominant contribution comes from low energy states
which overlap only in the classically forbidden region.
In this case therefore the reaction includes a nuclear tunnelling process
and requires the treatment of such quantum effects to be described adequately.

However, unless the internal states are known,
which would be the case for instance if the potential-energy surfaces are assumed to be harmonic,
this formulation cannot be applied straightforwardly.
In the present work, we seek a semiclassical approximation
which allows us to evaluate the rate in the golden-rule limit for complex systems
by avoiding the computation of the wave functions explicitly.

We introduce the Green's functions (fixed-energy propagators)
describing nuclear dynamics on each of the electronic states independently,
defined equivalently by the following two expressions
\begin{align}
	\op{G}_n(E) &= \lim_{\eta\rightarrow0^+} (E + \iu\eta - \op{H}_n)^{-1} %
\\
	&= \lim_{\eta\rightarrow0^+} - \frac{\iu}{\hbar} \int_0^\infty \eu{-\iu\op{H}_nt/\hbar} \, \eu{\iu(E+\iu\eta)t/\hbar} \, \rmd t.
	\label{Laplace}
\end{align}
The imaginary part is related
to the density of states
by
$\Im \op{G}_n(E) = - \pi \delta(E-\op{H}_n)$.
The reaction probability can thus be written as
\begin{multline} \label{Pgreens}
	P(E) = 4 \iint \Delta(\mat{x}') \Im \braket{\mat{x}'|\op{G}_0(E)|\mat{x}''}
		\\ \times \Delta(\mat{x}'') \Im \braket{\mat{x}''|\op{G}_1(E)|\mat{x}'} \rmd\mat{x}' \rmd\mat{x}''.
\end{multline}
A similar formula was derived by Miller et al. \cite{Miller1983rate}
for adiabatic reactions
in which it was also noticed that only the imaginary part of the Green's function need be known
in order to obtain the reaction rate.

An exact evaluation of the Green's functions will be impossible for complex systems
as this is equivalent to a complete solution of the Schr\"odinger equation.
Nor does the microcanonical density operator treated in a path-integral representation \cite{Miller1975rate,Lawson2000microcanonical}
lead to a practical computational technique.
We therefore require a simpler semiclassical description of the imaginary part of the Green's functions
in the classically forbidden region.
The derivation of this is given in the following section.

\section{Semiclassical Green's functions}
\label{sec:greens}

Like the wave function
which solves the Schr\"odinger equation for the Hamiltonian in \eqn{Hn},
the Green's function defined by \eqn{Laplace}
contains all information required to study the nuclear dynamics.
It would therefore be a very useful object to obtain
and apply to a wide range of problems,
although it is in general as difficult to compute exactly as the wave function itself.
However, it can easily be defined using Feynman's path-integral propagator, \cite{Feynman}
from which one can take a semiclassical approximation replacing the path integral as a sum over classical trajectories.

In many previous applications, semiclassical Green's functions
were employed to describe quantization
in bound states. \cite{Gutzwiller1971orbits,Miller1972quantization}
This required locating periodic orbits in the classically allowed region,
for which there may be numerous possibilities,
many of which are unstable,
especially in large complex systems exhibiting chaotic dynamics.
However, in this work, the most important quantum effect is that of tunnelling
and we neglect quantization in the reactant and product potential wells.
This means that we are interested in evaluating the Green's function only in the classically forbidden region.
As we shall show, this requires us to locate a single trajectory, 
and is therefore expected to lead to computationally feasible methods even in complex systems.

Gutzwiller
\cite{Gutzwiller1967semiclassical,Gutzwiller1971orbits,GutzwillerBook}
has derived a semiclassical approximation to the Green's function, $\braket{\mat{x}'|\op{G}_n(E)|\mat{x}''}$,
in the classically allowed region,
i.e.\ where $V_n(\mat{x}') < E$ and $V_n(\mat{x}'') < E$.
The derivation starts from
the van-Vleck propagator,
a semiclassical approximation to $\braket{\mat{x}'|\eu{-\iu\op{H}_n(t'-t'')/\hbar}|\mat{x}''}$, \cite{Gutzwiller1967semiclassical}
\begin{align}
	K_n(\mat{x}',\mat{x}'',t'-t'') &= \sum_\text{cl. traj.} \frac{\sqrt{C_n}}{(2\pi\iu\hbar)^{f/2}} \, \eu{\iu S_n/\hbar} %
	\label{vanVleck}
\\
	C_n &= \left| - \pders{S_n}{\mat{x}'}{\mat{x}''} \right|,
\end{align}
where the full action is
\begin{align} \label{Sn}
	S_n &\equiv S_n(\mat{x}',\mat{x}'',t'-t'') \nonumber\\
		&= \int_{t''}^{t'} \left[ \half m \left|\pder{\mat{x}(t)}{t}\right|^2 - V_n\big(\mat{x}(t)\big) \right] \rmd t.
\end{align}
Here, $\mat{x}(t)$ is a classical trajectory which travels from $\mat{x}''$ at time $t''$ to $\mat{x}'$ at time $t'$,
and we sum over all such trajectories.
To avoid clutter,
the transpose is implied on the second partial derivative variable of Hessian matrices throughout,
e.g.\ $\pders{S_n}{\mat{x}'}{(\mat{x}'')^\T}$.
As the Hamiltonian is time-independent, we can without loss of generality set $t''=0$.
We note that $-\pder{S_n}{t'}$ gives the energy conserved along the trajectory.

Using van-Vleck's propagator in \eqn{Laplace},
Gutzwiller employed a stationary-phase evaluation of the Laplace integral.
The stationary-phase points solve
\begin{align}
	\pder{}{t_n}S_n(\mat{x}',\mat{x}'',t_n) + E = 0,
	\label{stationaryphase}
\end{align}
which picks out times, $t_n$, defining trajectories with the required energy.
This gives a semiclassical approximation for the Green's function
as a sum over classical trajectories with energy $E$, \cite{GutzwillerBook,Kleinert}
\begin{align}
	\label{Gcl}
	G_n(\mat{x}',\mat{x}'',E)
	=
	\sum_\text{cl. traj.}
			\frac{2\pi \sqrt{|D_n|}}{(2\pi\iu\hbar)^{(f+1)/2}}
			\, \eu{\iu W_n/\hbar - \iu\nu\pi/2}.
\end{align}
Now the same classical trajectories can
are formally defined %
as continuous paths
starting at $\mat{x}''$ and ending at $\mat{x}'$
and giving a stationary value of the abbreviated action, \cite{Whittaker,Goldstein}
\begin{align}
	W_n &\equiv W_n(\mat{x}',\mat{x}'',E) = \int_{\mat{x}(q)=\mat{x}''}^{\mat{x}(q)=\mat{x}'} p_n \, \rmd q,
\end{align}
which is a line integral along the trajectory $\mat{x}(q)$ represented by the generalized coordinate $q$.
The magnitude of the momentum at any point along the trajectory is
\begin{align}
	p_n &= \sqrt{2m[E-V_n(\mat{x})]}.
	\label{p}
\end{align}

The prefactor is found from
the determinant of an \mbox{$(f+1)$}-dimensional square matrix, %
\begin{align}
	D_n &= (-1)^{f+1} \left| \begin{matrix} \frac{\partial^2 W_n}{\partial \mat{x}' \partial \mat{x}''} &
										\frac{\partial^2 W_n}{\partial \mat{x}'\partial E } \\
										\frac{\partial^2 W_n}{\partial E \partial \mat{x}''} &
										\frac{\partial^2 W_n}{\partial E^2} \end{matrix} \right|.
\end{align}
Following Gutzwiller \cite{Gutzwiller1967semiclassical},
we take the absolute value of this determinant and introduce a phase term in \eqn{Gcl} determined by the Maslov-Morse index, $\nu$,
which is given by the number of conjugate points along the trajectory.
For our purposes, conjugate points occur where the trajectory
encounters a turning point, where $V_n(\mat{x})=E$,
and comes instantaneously to rest. \cite{GutzwillerBook}
We call this a bounce.

We now derive an equivalent semiclassical Green's function formalism for the classically forbidden region,
where $V_n(\mat{x}') > E$ and $V_n(\mat{x}'') > E$.
Some work in this direction has already been completed
to describe tunnelling using imaginary-time trajectories, 
using similar approaches to that outlined in this paper.
Some generalization is however required
as results in the literature 
considered only pathways which pass through the forbidden region but with end points outside,
\cite{Freed1972semiclassical,Miller1979periodic}
or treated only certain one-dimensional systems.
\cite{Holstein1982tunnelling,Carlitz1985semiclassical}
We also introduce a transition-state theory approximation below
which allows us 
to present a simple expression for the Green's function between classically forbidden end points
in the general multidimensional case.

Using the semiclassical van-Vleck propagator in \eqn{Laplace} 
gives stationary-phase points which again solve \eqn{stationaryphase}.
As before, solutions correspond to classical trajectories travelling from $\mat{x}''$ to $\mat{x}'$ in a given time.
However, classical dynamics in the forbidden region is only possible in imaginary time,
as it is well-known that this
is equivalent to real-time dynamics on an upside-down potential-energy surface. \cite{Miller1971density}
Therefore, whereas in the classically allowed region these stationary points lay on the real-time axis,
for the forbidden region they are on the imaginary-time axis.
The action for these trajectories is purely imaginary.

Let us consider the nature of these imaginary-time trajectories.
In the $f$-dimensional coordinate space,
there will be a $(f-1)$-dimensional ``turning'' surface  on which $V_n(\mat{x})=E$.
This surface is a generalization of the turning point, which separates the allowed from the forbidden region,
and imaginary-time trajectories can bounce off it.
Possible solutions therefore correspond to trajectories which travel directly from one end point to the other
or those which bounce of the surface in-between.
Because the dynamics are time reversible,
a bouncing trajectory will always travel along the same pathway before and after the bounce,
which always occurs normal to the turning surface.
The direct trajectory corresponds to $t_\text{d}=-\iu\tau_\text{d}$,
and the bounce to $t_\text{b}=-\iu\tau_\text{b}$,
where $\tau_\text{b}>\tau_\text{d}>0$.
The stationary point of the former is a maximum along the imaginary axis but a minimum along the real axis,
whereas the latter is vice versa.
Note that the reverse of these trajectories also exist with $\Im t>0$
but occur in the wrong part of the complex plane to be of interest for the contour integration which we shall perform.

\begin{figure}
	\includegraphics{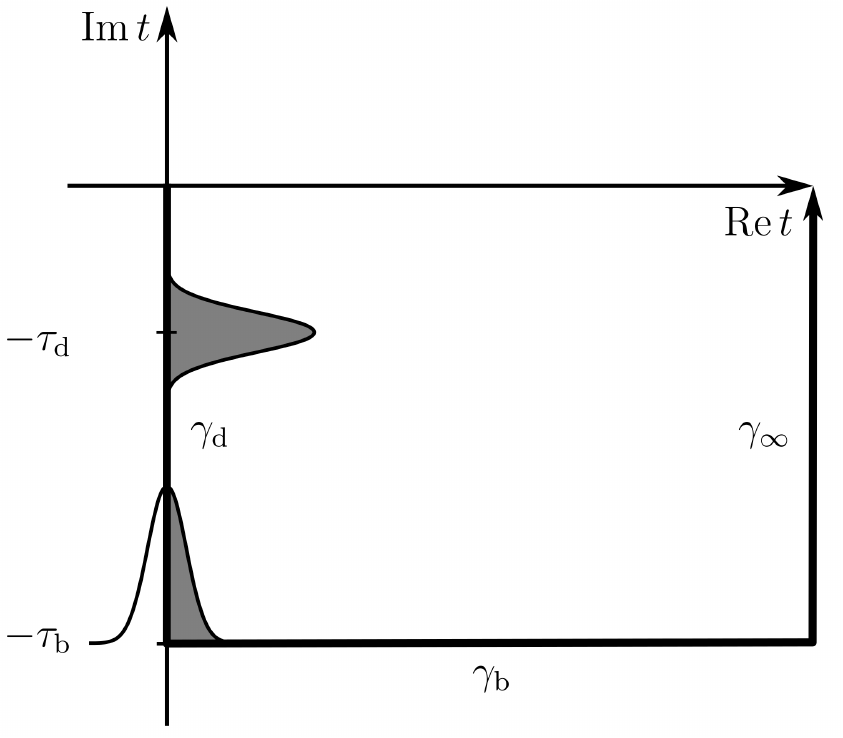}
	\caption{Representation of the deformed contour for the Laplace integral in the classically forbidden region.
	The two marked stationary-phase points correspond to the direct and bounce trajectories.}
	\label{fig:contour}
\end{figure}

For the simplest cases of tunnelling into an unbound potential,
such as the one-dimensional linear $V_n(x)=\kappa_n x$,
only these two classical trajectories can exist at a given energy.
However, a simple one-dimensional bound potential,
such as the harmonic oscillator,
also supports trajectories %
which pass through the turning surface into the classically allowed region
where it bounces any odd number of times,
before returning to the forbidden region. \cite{Carlitz1985semiclassical}
These correspond to complex times with $\Re t>0$ %
and contribute to the real part of the action $S_n(\mat{x}',\mat{x}'',t)$
and hence to the phase of the Green's function.
However, we are ultimately interested in condensed-phase systems where the potentials are multidimensional
and give rise to chaotic dynamics. \cite{GutzwillerBook}
This causes the phases to mostly cancel each other out when we take further integrals over the Green's functions.
We shall therefore assume that we can ignore the stationary points with $\Re t>0$
and compute a steepest-descent integration along the contour $\gamma_\text{d}+\gamma_\text{b}+\gamma_\infty$ shown in \fig{contour}
such that it passes through the two stationary points.
A small positive value of $\eta$
causes the function to decay with $\Re t$
and ensures that the integral along $\gamma_\infty$ is 0.

In this way,
we are restricting the information needed about the system to the classically forbidden region.
Our approximation is a form of quantum TST
as we are effectively ignoring coherence effects at long times.
The coherences are responsible for the discrete energy levels of a finite bound system, \cite{Gutzwiller1971orbits,Miller1972quantization}
and the zero-point energy of an infinite system. \cite{Weiss}
In many condensed-phase systems, these effects do not play an important role and can be safely ignored.
Only at very low energies will we find a serious error due to the discrete nature of states in the wells
which cannot be described using this approach.

More complex potential forms may support many bounce or direct trajectories
as well as those which bounce more than once.
In such cases, one can either sum over all possibilities
or turn to a path-integral sampling scheme which avoids the steepest-descent integration in position variables.
In this work, we derive only a semiclassical steepest-descent rate theory
and shall assume the simplest case that only one trajectory of each type exists,
but still without specifying the exact form of the surface.
We shall show in Paper II \cite{GoldenRPI} how Wolynes' quantum instanton method, \cite{Wolynes1987nonadiabatic,Bader1990golden,Zheng1989ET,*Zheng1991ET}
which can be used for systems with rough potentials,
is related to our golden-rule instanton approach.

We can then follow a very similar derivation to that of Gutzwiller's in the allowed region
performing steepest-descent integrations about the points $t_\text{d/b}$.
One must take particular note that our integration along $\gamma_\text{b}$ passes over only half of the Gaussian peak centred at $t_\text{b}$.
This gives the following semiclassical approximation to the Green's function:
\begin{align} \label{G}
	\bar{G}_n(\mat{x}',\mat{x}'',E)
	= \Gamma_n^\text{d}(\mat{x}',\mat{x}'',E) + \half \Gamma_n^\text{b}(\mat{x}',\mat{x}'',E),
\end{align}
where the functions $\Gamma_n^\text{d/b}$ are defined either by the direct or bounce imaginary-time trajectory as appropriate.
These trajectories are continuous paths
starting at $\mat{x}''$ and ending at $\mat{x}'$
which give a stationary value of the imaginary abbreviated action,
\begin{align}
	\label{barW}
	\bar{W}_n &\equiv \bar{W}_n(\mat{x}',\mat{x}'',E) = \int_{\mat{x}(q)=\mat{x}''}^{\mat{x}(q)=\mat{x}'} \bar{p}_n \, \rmd q \\
	\bar{p}_n &= \sqrt{2m[V_n(\mat{x})-E]}
	\label{barp}
\\
	\Gamma_n^\text{d/b}(\mat{x}',\mat{x}'',E) &= %
			- \frac{2\pi\sqrt{|\bar{D}_n|}}{(2\pi\hbar)^{(f+1)/2}}
			\, \eu{-\bar{W}_n/\hbar - \iu\nu\pi/2} \\
	\bar{D}_n &= (-1)^{f+1} \left| \begin{matrix} \frac{\partial^2 \bar{W}_n}{\partial \mat{x}' \partial \mat{x}''} &
										\frac{\partial^2 \bar{W}_n}{\partial \mat{x}'\partial E } \\
										\frac{\partial^2 \bar{W}_n}{\partial E \partial \mat{x}''} &
										\frac{\partial^2 \bar{W}_n}{\partial E^2} \end{matrix} \right|.
\end{align}
We use the notation of a bar over all variables related to imaginary-time propagation.
Note that the imaginary action $\bar{W}$ is so called because it is defined as $-\iu W$.
In this region, like $\bar{p}_n=-\iu p_n$, it is always real and positive.
Unlike the Green's functions in the classically allowed region, which were oscillatory,
these decay exponentially with $\bar{W}$.

The direct trajectory, with $\nu=0$, therefore contributes to the real part,
and the bounce, with $\nu=1$, to the imaginary part of the semiclassical Green's function.
Even if trajectories exist with more than one bounce,
they can normally be ignored
as their large actions will ensure that they do not dominate either the real or imaginary parts.

The semiclassical formulae
suffer in the same way as the
Wentzel-Kramers-Brillouin (WKB) approximation \cite{Schiff}
from poles at the turning points of a trajectory.
That is, at a turning point where \mbox{$V_n(\mat{x}) = E$}, the prefactor $\bar{D}_n$ goes to infinity,
as can be seen by transforming to a coordinate basis parallel and perpendicular to the trajectory. \cite{Gutzwiller1971orbits}
However, we have found them to be a good approximation
to the exact Green's functions
when $\mat{x}'$ and $\mat{x}''$ are far from the turning points,
which as we shall show is all that is required for a steepest-descent evaluation of the golden-rule rate constant.

We have shown that the Green's function is not just a simple analytic continuation of Gutzwiller's formula
because a factor of a half appears in the imaginary part.
This is in agreement with previous work \cite{Holstein1982tunnelling,Carlitz1985semiclassical}
which explain the one-dimensional WKB connection formulae. \cite{Schiff}
Interestingly the factor also appears in the \ImF\ approach 
where it is argued that only half of the imaginary barrier partition function is required. \cite{Uses_of_Instantons,Kleinert}
The derivation given here based on
the contour integral used to compute the Laplace transform of the van-Vleck propagator
is more direct and requires no analytic continuation.

For illustration,
we shall give a simple example of the semiclassical Green's function
for the one-dimensional linear potential-energy surface \mbox{$V_n(x)=\kappa_n x$}.
The semiclassical approximation in the classically allowed region, $\kappa_n x',\kappa_n x''<E$, is given by
\begin{align}
	G_n(x',x'',E)
		&= - \frac{1}{\hbar}\sqrt\frac{m^2}{p_n(x')p_n(x'')} \left[\iu\eu{\iu W_n^\text{d}/\hbar} + \eu{\iu W_n^\text{b}/\hbar}\right],
	\label{Gapprox}
\end{align}
with
\begin{subequations} \label{Wlinear}
\begin{align}
	W_n^\text{d} &= \left| \frac{p_n(x'')^3-p_n(x')^3}{3m\kappa_n} \right|
\\
	W_n^\text{b} &= \frac{p_n(x'')^3+p_n(x')^3}{|3m\kappa_n|},
\end{align}
\end{subequations}
or in the forbidden region,
$\kappa_n x',\kappa_n x''>E$,
\begin{align}
	\bar{G}_n(x',x'',E)
		&= - \frac{1}{\hbar}\sqrt\frac{m^2}{\bar{p}_n(x')\bar{p}_n(x'')} \left[ \eu{-\bar{W}_n^\text{d}/\hbar} + \frac{\iu}{2}\eu{-\bar{W}_n^\text{b}/\hbar} \right],
	\label{Gbarapprox}
\end{align}
with $\bar{W}_n^\text{d/b}$ defined as \eqs{Wlinear} with bars added over the variables
and momenta defined as in \eqn{p} and \eqn{barp}.
This is the same semiclassical result as found in \Ref{Carlitz1985semiclassical}.

The wave functions for this linear potential are known to be
$\psi_n(x;E) = \sqrt{\alpha} \Ai\left(z(x;E)\right)$,
where $\alpha = |4m^2/\hbar^4\kappa_n|^{1/3}$
and $z(x;E) = (2m/\kappa_n^2\hbar^2)^{1/3} (\kappa_n x - E)$.
The exact Green's function for this linear potential can be found using the procedure outlined in \Ref{Bakhrakh1971Greens}
from the Wronskian, $w$. %
It is
\begin{align}
	\braket{x'|\op{G}_n(E)|x''} &= \frac{2 m \alpha}{\hbar^2w} \Ai\left(z_>\right) \Ai\left(\eu{2\iu\pi/3}z_< \right)
	\label{Glinear}
\\
	w &= m \frac{-1 + \iu/\sqrt{3}}{\hbar^{2/3} \, \Gamma\!\left(\tfrac{1}{3}\right) \Gamma\!\left(\tfrac{2}{3}\right)},
\end{align}
where $z_<$ and $z_>$ are the lesser and greater of $z(x';E)$ and $z(x'';E)$.

\begin{figure}
	\includegraphics{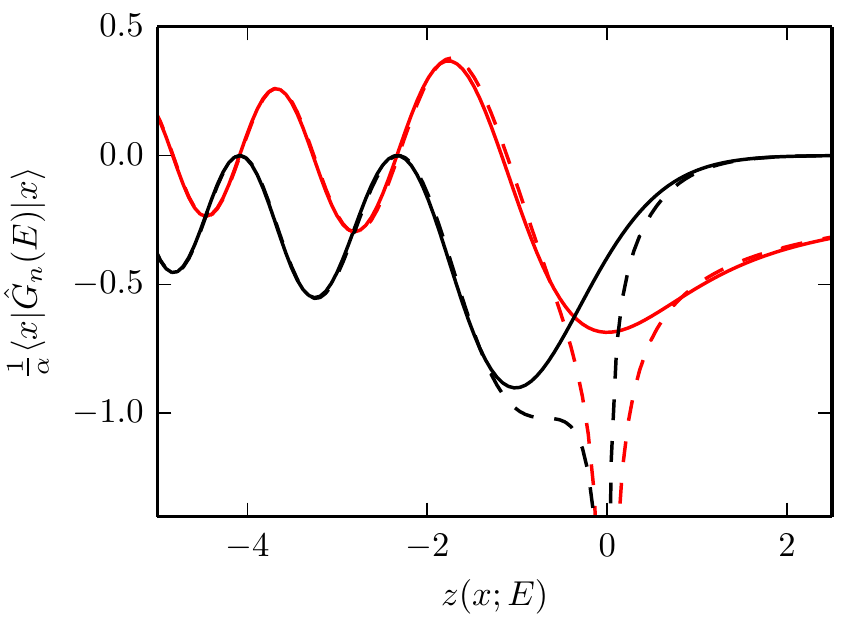}
	\caption{The diagonal elements of the Green's function for a linear potential
	and plotted with real and imaginary parts in red and black
	against a dimensionless function of position and energy.
	The exact form, \eqn{Glinear}, is shown with solid lines
	compared to the semiclassical approximations, Eqs. \eqref{Gapprox} and \eqref{Gbarapprox}, with dashed lines.}
	\label{fig:airy}
\end{figure}

The semiclassical approximation is compared to the exact Green's functions in \fig{airy},
showing good agreement everywhere except near the classical turning point $z=0$.
In fact the imaginary part of the semiclassical Green's function approaches the exact result asymptotically
as can be proved using
$\Im\braket{x'|\op{G}_n(E)|x''} = - \pi \psi_n(x';E) \psi_n(x'';E)$
and an asymptotic approximation for the Airy function, \cite{Abramowitz}
accurate for $|z|\rightarrow\infty$
given by,
\begin{equation} \label{approxAi}
	\Ai(z) \approx
		\begin{cases} \frac{\exp\left[-\frac{2}{3}z^{3/2}\right]}{2\sqrt{\pi}z^{1/4}} & z > 0 \\
						  \frac{\sin\left[\frac{2}{3}(-z)^{3/2} + \frac{\pi}{4}\right]}{\sqrt{\pi}(-z)^{1/4}} & z < 0.
		\end{cases}
\end{equation}
Note again that the factor of half appearing in \eqn{G} was necessary for this equivalence.

\section{Golden-Rule Instanton Formulation}
\label{sec:instanton}

In this section, we shall derive an approximate formula for the golden-rule rate
using two semiclassical Green's functions in the classically forbidden region.
We shall perform this derivation
using a steepest-descent integral first over the positions
and then over energy in the exact expression for the rate.
This picks out two imaginary-time classical trajectories,
which when joined together are known as the instanton.

The usual procedure for performing steepest-descent integrals
assumes that pre-exponential terms are approximately constant over the range in which the exponential dominates.
That is for functions $A(\mat{q})$ and $B(\mat{q})$ of a $d$-dimensional variable $\mat{q}$,
\begin{align}
	\int_\text{SD} B(\mat{q}) \, \eu{-A(\mat{q})/\hbar} \, \rmd \mat{q}
	=
	(2\pi\hbar)^{d/2} B(\mat{q}^\ddag) \left|\frac{\partial^2A}{\partial\mat{q}\partial\mat{q}}\right|_{\mat{q}=\mat{q}^\ddag}^{-\half} \eu{-A(\mat{q}^\ddag)/\hbar},
\end{align}
where $\mat{q}^\ddag$ is defined such that $A(\mat{q}^\ddag)$ is a minimum. %
This is also known as the semiclassical approximation because,
as long as $B(\mat{q}^\ddag)\neq0$,
it gives the term with lowest order of $\hbar$ correctly.
The error in the approximation is always an order of $\hbar$ higher
and thus becomes exact if $\hbar\rightarrow0$.

This steepest-descent approach
requires that only one minimum of the function $A(\mat{q})$ exists,
but is easily extended to treat multiple \textit{well-separated} minima
by summing over the contributions.
However, like other instanton approaches, \cite{Miller1975semiclassical,Benderskii,rpinst}
the resulting formulation will not be able to treat the rough potential surfaces found for reactions in liquids
where many local minima exist whose steepest-descent integrands would overlap.
Such problems are better treated using path-integral Monte Carlo or molecular dynamics approaches, \cite{Wolynes1987nonadiabatic,Ceperley1995PathIntegrals,Marx1996PIMD}
which we discuss in the following paper. \cite{GoldenRPI}
The instanton approach derived here is only directly applicable to solid \cite{EsquinaziBook}
or gas-phase systems as well as system-bath models of condensed-phase electron transfer. \cite{Weiss}
However, it is the derivation and physical interpretation of a rate formula which is the focus of the present work,
and so we will assume for now that the potentials are sufficiently smooth.

As already pointed out above, for the calculation of electron-transfer rates
it is the classically forbidden region
which dominates the result.
We note that the Green's function will be much simpler to work with numerically
in this region,
where except for phases originating from the conjugate points,
there is a real exponent,
and it is thus not an oscillatory function.
By a similar principle other imaginary-time path-integral calculations \cite{Chandler+Wolynes1981}
are numerically tractable,
including, for instance, instanton calculations of adiabatic reaction rates %
and ring-polymer transition-state theory (RPTST). \cite{rpinst,Hele2013QTST}

In \secref{Lagrangian} %
we transform our rate expression from the language of the Hamilton-Jacobi formalism,
where the trajectories are defined by their energy,
to the language of Lagrangian dynamics,
where they are defined by the elapsed time.

As our formula for the rate will be derived from a steepest-descent integration over the position coordinates,
a consistent calculation of the reactant partition function is
\begin{align}
	Z_0 \approx \prod_{j=1}^f \frac{1}{2\sinh{\half\beta\hbar\omega_j}},
	\label{Z0}
\end{align}
where $\omega_j$ are the normal-mode frequencies at the minimum of $V_0(\mat{x})$.
This form assumes that there are no translation or rotational degrees of freedom but can be easily modified to
treat gas-phase problems 
by projecting out such modes in the usual manner.
In this case the steepest-descent approximation is equivalent to a local harmonic approximation for the region near the bottom of the reactant well.
However, this is only a minor approximation,
and in what follows, we do not assume that this harmonic approximation holds for the whole reactant potential.

\subsection{Hamilton-Jacobi formalism}
\label{sec:HJ}

\begin{figure}
	\includegraphics{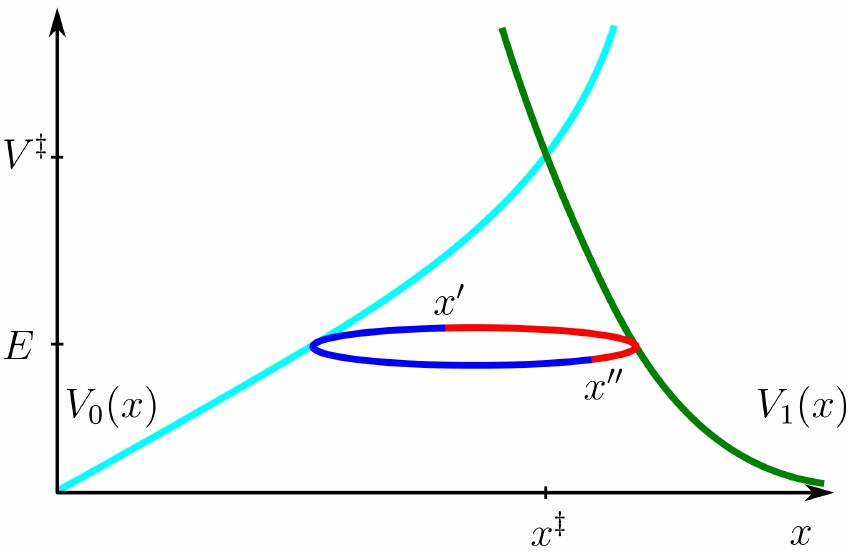}
	\caption{Schematic showing the two imaginary-time bounce trajectories with energy $E$
	near the crossing point in a one-dimensional, two-state system.
	The blue trajectory is on $\ket{0}$ and the red on $\ket{1}$.
	The steepest-descent integration of positions will be taken about the crossing point $x'=x''=x^\ddag$
	at which $V_0(x^\ddag)=V_1(x^\ddag)=V^\ddag$.
	}
	\label{fig:oneD}
\end{figure}

We insert our semiclassical approximation for the Green's functions %
into \eqn{Pgreens}.
This requires only the imaginary parts
which are, according to our approximate formula, \eqn{G},
given by trajectories with a conjugate point,
i.e.\ those which bounce.
Note also that the energies of the Green's functions on each electronic state are equal.
We thus obtain the semiclassical reaction probability
\begin{align} \label{SCRP}
	P_\text{SC}(E) &= \iint_\text{SD} \frac{\Delta(\mat{x}')\Delta(\mat{x}'')}{\hbar^2} \frac{\sqrt{\bar{D}_0\bar{D}_1}}{(2\pi\hbar)^{f-1}} \, \eu{-\bar{W}/\hbar} \, \rmd\mat{x}' \rmd\mat{x}'',
\end{align}
where the combined action is $\bar{W}=\bar{W}_0+\bar{W}_1$
and the trajectories considered are those which bounce once between the end points $\mat{x}'$ and $\mat{x}''$
as shown in \fig{oneD}.
We have assumed that there is only one such trajectory on each surface.
If more than one exists, one should either take only the dominant trajectory with the smallest value of $\bar{W}$
or sum over all possibilities.
Note that the integral in \eqn{SCRP} shall be computed using the method of steepest descent
which avoids including the spurious poles of $\bar{D}_n$.

The steepest-descent integration is taken about the point
$\mat{x}'=\mat{x}''=\mat{x}^\ddag$,
at which
\begin{subequations}
\label{pderW}
\begin{align}
	\pder{\bar{W}}{\mat{x}'} &= \bar{\mat{p}}_0' - \bar{\mat{p}}_1' = \mat{0} \\
	\pder{\bar{W}}{\mat{x}''} &= -\bar{\mat{p}}_0'' + \bar{\mat{p}}_1'' = \mat{0},
\end{align}
\end{subequations}
where $\bar{\mat{p}}_n'$ (or its equivalent with double primes)
is the imaginary momentum of the trajectory on surface $n$ at $\mat{x}'$;
it has magnitude $\bar{p}_n(\mat{x}')$ and direction pointing along the trajectory.
As classical mechanics is time-reversible, the direction chosen is immaterial, but should be consistent.

The fact that the energy $E$ is equal for both trajectories
by construction
implies that $x^\ddag$ lies on the crossing seam.
This has physical significance showing that although
the bouncing instanton trajectory is delocalized,
the hop between electronic states occurs dominantly on the crossing seam.
For a one-dimensional system this point is uniquely defined,
but in general, the seam is an $(f-1)$-dimensional surface
and the hopping point $\mat{x}^\ddag$ varies for trajectories of different energy.
Because the imaginary momenta on each surface cancel at the hopping point,
the trajectories must join smoothly into each other to form a periodic orbit.
This, along with the constraint that both trajectories must reach a turning point
defines the hopping point $\mat{x}^\ddag$ for energy $E$.
Note however that this does not imply that the momentum at this point is necessarily normal to the $V_0(\mat{x})=V_1(\mat{x})$ surface.

In the following, all terms
shall be evaluated at this hopping point $\mat{x}'=\mat{x}''=\mat{x}^\ddag$.
This includes the electronic coupling %
for which we therefore need only one value, denoted \mbox{$\Delta = \Delta(\mat{x}^\ddag)$}.
The semiclassical reaction probability is thus defined as
\begin{align}
	P_\text{SC}(E)
	&= \frac{2\pi}{\hbar} \Delta^2
			\sqrt{\bar{D}_0 \bar{D}_1}
			\left| \begin{matrix}
										\frac{\partial^2 \bar{W}}{\partial\mat{x}' \partial\mat{x}'} &
										\frac{\partial^2 \bar{W}}{\partial\mat{x}'\partial\mat{x}''} \\
										\frac{\partial^2 \bar{W}}{\partial\mat{x}''\partial\mat{x}'} &
										\frac{\partial^2 \bar{W}}{\partial\mat{x}''\partial\mat{x}''} \end{matrix}
								\right|^{-\half} %
			\eu{-\bar{W}/\hbar}.
	\label{PSC}
\end{align}

This is the form given for a system in the so-called normal regime
where the reactant and product minima lie on opposite sides of the crossing seam.
When they lie on the same side, known as the Marcus inverted regime,
it is clear that a different Ansatz would be required to define the instanton
because \eqs{pderW} no longer define the stationary point.

The result in \eqn{PSC} could be used directly
to compute microcanonical reaction rates. \cite{Miller1993QTST}
The approximation is only valid for energies lower than the activation energy $V^\ddag$
and will also deviate strongly from the exact result at very low energies in condensed-phase systems.
This is due to the transition-state theory assumption
which ignores the true density of states in the reactant well
and neglects its zero-point energy.
We concentrate however on computing the thermal rate which is found by
integration over all energies weighted by the Boltzmann distribution.
It would also be possible to extend the theory to use other energy distributions and describe certain nonequilibrium effects.

Inserting the approximation for the reaction probability into \eqn{kthermal}
and performing a steepest descent integral over $E$
gives the semiclassical result that we seek:
\begin{multline}
	k_\text{SC} Z_0
	= \sqrt{2\pi\hbar} \, \frac{\Delta^2}{\hbar^2}
			\sqrt{\bar{D}_0 \bar{D}_1}
			\left| \begin{matrix}
										\frac{\partial^2 \bar{W}}{\partial\mat{x}' \partial\mat{x}'} &
										\frac{\partial^2 \bar{W}}{\partial\mat{x}'\partial\mat{x}''} \\
										\frac{\partial^2 \bar{W}}{\partial\mat{x}''\partial\mat{x}'} &
										\frac{\partial^2 \bar{W}}{\partial\mat{x}''\partial\mat{x}''} \end{matrix}
								\right|^{-\half} %
		\\\times
			\left(\frac{\rmd^2 \bar{W}}{\rmd E^2}\right)^{-\half}
			\eu{-\bar{W}/\hbar - \beta E},
	\label{kWfull}
\end{multline}
where the value is given at energy $E$ which solves
$\frac{\partial\bar{W}}{\partial E}+\beta\hbar=0$ for a given temperature.
The full derivative implies that $\mat{x}'$ and $\mat{x}''$ are not held fixed
but are moved to the appropriate value of $\mat{x}^\ddag$,
which is the stationary point of
$\bar{W}$ for each given energy.
Of course in one-dimensional systems where the hopping point is always the same,
there is no difference here between the full and partial derivative with respect to $E$.

\begin{figure}
	\includegraphics{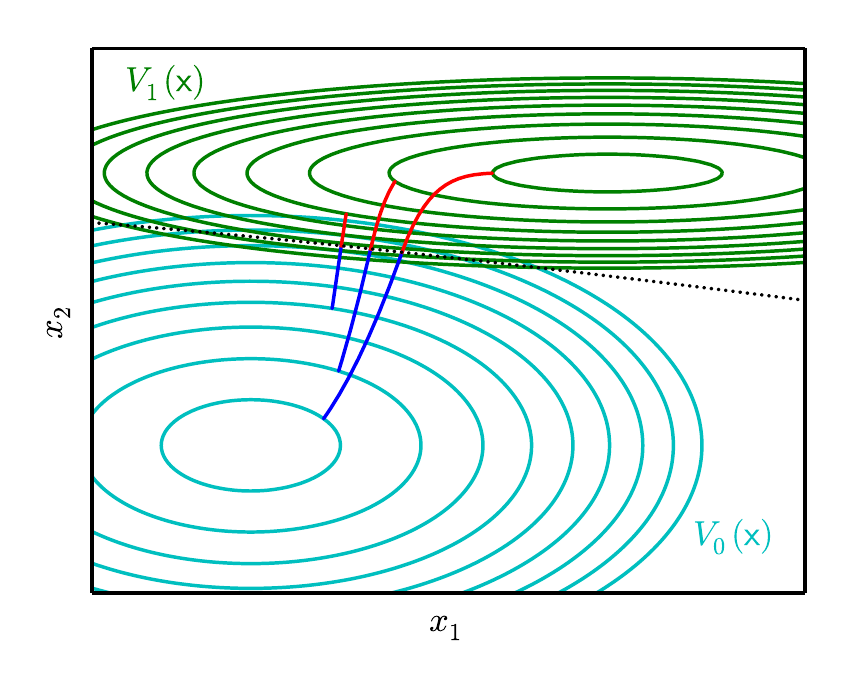}
	\caption{Schematic showing three instanton paths (blue and red) in a two-dimensional system at different temperatures.
	The shortest path corresponds to the highest temperature which tunnels at high energy,
	whereas at low temperature, the tunnelling pathway is longer and at low energy.
	Contours are shown for both potential-energy surfaces
	such that one contour level is chosen to be the energy $E$ of each of the instantons.
	In each case, this contour level surrounds the classically allowed region where the imaginary-time trajectory cannot enter
	and the instanton bounces normal to this surface.
	The hop occurs at $\mat{x}^\ddag$ without affecting the momentum of the orbit
	and is represented by the change in colour of the pathway.
	These points vary with temperature but are always located on
	the crossing seam $V_0(\mat{x})=V_1(\mat{x})$, shown by the dotted line.
	}
	\label{fig:twoD}
\end{figure}

Because $\tau_n=-\frac{\partial\bar{W}_n}{\partial E}$ gives the imaginary time of the trajectory,
the energy is chosen by the steepest-descent integration is that which ensures that the total imaginary time taken by the orbit
is $\tau_0+\tau_1=\beta\hbar$.
We have therefore obtained a formula similar to 
the semiclassical description
of the quantum Boltzmann distribution, \cite{Miller1971density}
which %
also leads to the usual derivations of adiabatic instantons
in terms of imaginary-time periodic orbits. \cite{Miller1975semiclassical,rpinst}
The golden-rule instanton is thus a periodic orbit of constant energy and total imaginary time $\beta\hbar$.
It follows a continuous path on the $V_0(\mat{x})$ surface from $\mat{x}^\ddag$ to a turning point
before retracing its steps back to $\mat{x}^\ddag$.
Here it hops to the $V_1(\mat{x})$ surface, without needing to modify its momentum as the potentials are equal,
and performs a similar bounce before returning.
The periodic orbit must retrace itself after the bounce
because it must approach and depart from a turning point, at which it comes to rest, along the potential gradient.
In \fig{twoD}, we show how the periodic orbit and the hopping point $\mat{x}^\ddag$ are affected by the temperature parameter $\beta$.
Higher temperatures lead to shorter instantons
and thus a less delocalized and more ``classical'' reaction where tunnelling is less pronounced.
Lower temperatures allow more freedom for the instanton pathway to become curved,
an effect known as corner-cutting. \cite{Marcus1977cornercut,Benderskii}

\Eqn{kWfull} is only one of many possible ways to formulate the golden-rule instanton rate
and we consider now transforming it to an alternative form.
Applying the chain rule at the stationary point defined by \eqs{pderW} gives
\begin{align}
	\frac{\rmd^2\bar{W}}{\rmd E^2}
	&= \frac{\partial^2\bar{W}}{\partial E^2}
		+ \frac{\partial^2\bar{W}}{\partial E\partial\mat{x}'} \frac{\rmd\mat{x}'}{\rmd E}
	 	+ \frac{\partial^2\bar{W}}{\partial E\partial\mat{x}''} \frac{\rmd\mat{x}''}{\rmd E}
\end{align}
and %
\begin{align}
	\begin{pmatrix} \mat{0} \\ \mat{0} \end{pmatrix}
	= 
	\frac{\rmd}{\rmd E}
	\begin{pmatrix} \frac{\partial\bar{W}}{\partial\mat{x}'} \\ \frac{\partial\bar{W}}{\partial\mat{x}''} \end{pmatrix}
	=
	\begin{pmatrix} \frac{\partial^2\bar{W}}{\partial \mat{x}' \partial E} \\ \frac{\partial^2\bar{W}}{\partial\mat{x}'' \partial E} \end{pmatrix}
	+
	\begin{pmatrix} \frac{\partial^2\bar{W}}{\partial \mat{x}'\partial\mat{x}'} & \frac{\partial^2\bar{W}}{\partial \mat{x}'\partial\mat{x}''} \\
						 \frac{\partial^2\bar{W}}{\partial \mat{x}''\partial\mat{x}'} & \frac{\partial^2\bar{W}}{\partial \mat{x}''\partial\mat{x}''}
	\end{pmatrix}
	\begin{pmatrix} \frac{\rmd\mat{x}'}{\rmd E} \\ \frac{\rmd\mat{x}''}{\rmd E} \end{pmatrix}.
\end{align}
\begin{widetext}
Solving these linear equations leads to the following relationship
between full and partial derivatives:
\begin{align} \label{d2WdE2}
	\frac{\rmd^2\bar{W}}{\rmd E^2}
	= \frac{\partial^2\bar{W}}{\partial E^2}
	-	\begin{pmatrix} \frac{\partial^2\bar{W}}{\partial E \partial\mat{x}'} & \frac{\partial^2\bar{W}}{\partial E \partial\mat{x}''} \end{pmatrix}
		\begin{pmatrix} \frac{\partial^2\bar{W}}{\partial \mat{x}'\partial\mat{x}'} & \frac{\partial^2\bar{W}}{\partial \mat{x}'\partial\mat{x}''} \\
							 \frac{\partial^2\bar{W}}{\partial \mat{x}''\partial\mat{x}'} & \frac{\partial^2\bar{W}}{\partial \mat{x}''\partial\mat{x}''}
		\end{pmatrix}^{-1}
		\begin{pmatrix} \frac{\partial^2\bar{W}}{\partial\mat{x}' \partial E} \\ \frac{\partial^2\bar{W}}{\partial\mat{x}'' \partial E} \end{pmatrix},
\end{align}
such that the semiclassical rate can be written equivalently as
\begin{align}
	k_\text{SC} Z_0
	&= 
			\sqrt{2\pi\hbar} \, \frac{\Delta^2}{\hbar^2}
			\sqrt{\bar{D}_0 \bar{D}_1}
			\left| \begin{matrix}
										\frac{\partial^2 \bar{W}}{\partial\mat{x}' \partial\mat{x}'} &
										\frac{\partial^2 \bar{W}}{\partial\mat{x}'\partial\mat{x}''} &
										\frac{\partial^2 \bar{W}}{\partial\mat{x}'\partial E} \\
										\frac{\partial^2 \bar{W}}{\partial\mat{x}''\partial\mat{x}'} &
										\frac{\partial^2 \bar{W}}{\partial\mat{x}''\partial\mat{x}''} &
										\frac{\partial^2 \bar{W}}{\partial\mat{x}''\partial E} \\
										\frac{\partial^2 \bar{W}}{\partial E \partial\mat{x}'} &
										\frac{\partial^2 \bar{W}}{\partial E \partial\mat{x}''} &
										\frac{\partial^2 \bar{W}}{\partial E^2}
										\end{matrix}
								\right|^{-\half} %
			\eu{-\bar{W}/\hbar - \beta E},
	\label{kWpartial}
\end{align}
which is of course what would have been found by performing the steepest-descent integral over positions and energy in one step.
\end{widetext}

\Eqn{kWfull}, and equivalently \eqref{kWpartial},
which give a semiclassical approximation to the golden-rule rate
is the main result of this work.
If the instanton trajectory can be located and derivatives of its action evaluated,
the expression can be applied directly to complex systems.
We describe numerical methods for doing this in Paper II
based on optimizing discrete pathways.

We consider our derivation to be simpler and more rigorous
than previous golden-rule instanton approaches based on an extension of the \ImF\ approach. \cite{Langer1969ImF}
This procedure was applied by Cao and Voth \cite{Cao1997nonadiabatic}
to describe nonadiabatic transitions,
although there appears to be no physical argument to explain the use of the \ImF\ premise in this case.
Here no saddle point was found in the spatial degrees of freedom
but instead the imaginary part came from an analytic continuation
of the divergent integral along an imaginary time coordinate.
We shall however find that the formulae they obtained is very similar to that derived in this work.
This shows more clearly that these instanton approaches are a form of transition-state theory (TST),
i.e.\ real-time dynamical effects are ignored.
This would be harder to understand from the \ImF\ approach where time does not appear.

Our derivation was performed using the Hamilton-Jacobi formulation
to define the classical trajectories,
which was the natural choice for treating trajectories which were required to have equal energies.
However, there also exists an alternative formulation of classical dynamics
using the Lagrangian picture,
i.e.\ with a given imaginary time rather than energy.
In the next section we perform a Legendre transformation 
to find an equivalent definition
of the semiclassical rate formula.

\subsection{Lagrangian formalism}
\label{sec:Lagrangian}

The Lagrangian formulation of classical mechanics is based on Hamilton's principle function, \eqn{Sn}.
The imaginary time-version of this gives
the Euclidean action \cite{Miller1971density}
\begin{align}
	\bar{S}_n \equiv
	\bar{S}_n(\mat{x}',\mat{x}'',\tau_n)
	&= \int_0^{\tau_n} \left[\half m \left|\pder{\mat{x}(\tau)}{\tau}\right|^2 + V_n\big(\mat{x}(\tau)\big) \right] \rmd \tau
	\label{Sbar}
\\
	&= \bar{W}_n(\mat{x}',\mat{x}'',E) + E \tau_n,
	\label{Legendre}
\end{align}
where %
the trajectory,
$\mat{x}(\tau)$,
travels through the classically forbidden region from $\mat{x}(0)=\mat{x}''$ to $\mat{x}(\tau_n)=\mat{x}'$
with energy $E=\pder{\bar{S}_n}{\tau_n}$. %
Because $\tau_0+\tau_1=\beta\hbar$, the exponent of \eqn{kWpartial} is simply
$\bar{S}=\bar{S}_0+\bar{S}_1$,
which is the total action of the periodic orbit.
Because the energies of the two trajectories are equal by construction,
our values for $\tau_n$ must also obey $\pder{\bar{S}_0}{\tau_0}=\pder{\bar{S}_1}{\tau_1}$
or $\pder{\bar{S}}{\tau}=0$
where $\tau_0=\beta\hbar-\tau$, $\tau_1=\tau$
and thus
$\pder{}{\tau}=\pder{}{\tau_1}-\pder{}{\tau_0}$.
The hopping points are as before $\mat{x}'=\mat{x}''=\mat{x}^\ddag$.
Note that such trajectories with equal end points are forced to bounce as long as $\tau_n$ is not zero,
i.e.\ $0<\tau<\beta\hbar$.

\Eqn{Legendre} is a Legendre transformation
and expresses the relationship between the two equivalent dynamical formalisms.
Following \Ref{Kleinert} and taking derivatives,
we obtain the relations
\begin{subequations} \label{WtoS}
\begin{align} 
	\frac{\partial^2\bar{W}_n}{\partial E^2} &= - \left(\frac{\partial^2 \bar{S}_n}{\partial\tau_n^2}\right)^{-1} \\
	\frac{\partial^2\bar{W}_n}{\partial\mat{x}'\partial E} &= \frac{\partial^2\bar{S}_n}{\partial\mat{x}'\partial\tau_n} \left(\frac{\partial^2 \bar{S}_n}{\partial\tau_n^2}\right)^{-1} \\
	\frac{\partial^2\bar{W}_n}{\partial\mat{x}'\partial\mat{x}''} &= 
	\frac{\partial^2\bar{S}_n}{\partial\mat{x}'\partial\mat{x}''}
	- \frac{\partial^2\bar{S}_n}{\partial\mat{x}'\partial\tau_n} \left(\frac{\partial^2\bar{S}_n}{\partial\tau_n^2}\right)^{-1} \frac{\partial^2\bar{S}_n}{\partial\tau_n\partial\mat{x}''},
\end{align}
\end{subequations}
and the equivalents with any exchange of $\mat{x}'$ and $\mat{x}''$.

Therefore, if the derivatives of $\bar{S}_n$ are known,
it is a simple matter to identify all derivatives of $\bar{W}_n$.
Using \eqs{WtoS} and standard properties of determinants, \eqn{detblock}, some simple but laborious algebra gives
\begin{widetext}
\begin{align}
	\bar{D}_n
		&= (-1)^{f+1}  \frac{\partial^2 \bar{W}_n}{\partial E^2} \left| \frac{\partial^2 \bar{W}_n}{\partial \mat{x}' \partial \mat{x}''}
								- \frac{\partial^2 \bar{W}_n}{\partial \mat{x}' \partial E} \left(\frac{\partial^2 \bar{W}_n}{\partial E^2}\right)^{-1} \frac{\partial^2 \bar{W}_n}{\partial E \partial \mat{x}''} \right| \\
		&= \left(\frac{\partial^2\bar{S}_n}{\partial\tau_n^2}\right)^{-1} \bar{C}_n
\\
	\bar{C}_n &= \left|-\frac{\partial^2\bar{S}_n}{\partial\mat{x}'\partial\mat{x}''}\right|
\end{align}
and
\begin{align}
		\begin{vmatrix}
			\frac{\partial^2 \bar{W}}{\partial\mat{x}' \partial\mat{x}'} &
			\frac{\partial^2 \bar{W}}{\partial\mat{x}'\partial\mat{x}''} &
			\frac{\partial^2 \bar{W}}{\partial\mat{x}'\partial E} \\
			\frac{\partial^2 \bar{W}}{\partial\mat{x}''\partial\mat{x}'} &
			\frac{\partial^2 \bar{W}}{\partial\mat{x}''\partial\mat{x}''} &
			\frac{\partial^2 \bar{W}}{\partial\mat{x}''\partial E} \\
			\frac{\partial^2 \bar{W}}{\partial E \partial\mat{x}'} &
			\frac{\partial^2 \bar{W}}{\partial E \partial\mat{x}''} &
			\frac{\partial^2 \bar{W}}{\partial E^2}
		\end{vmatrix}
		&=
			-\left(\frac{\partial^2 \bar{S}_0}{\partial \tau_0^2} \frac{\partial^2 \bar{S}_1}{\partial \tau_1^2}\right)^{-1}
			\Sigma,
\end{align}
where
\begin{align}
		\Sigma &=
			\left(\pder[2]{\bar{S}_0}{\tau_0} + \pder[2]{\bar{S}_1}{\tau_1}\right)
			\left| \sum_{n,n'=0}^1 
			\begin{pmatrix}
				\frac{\partial^2 \bar{S}_n}{\partial\mat{x}' \partial\mat{x}'} &
				\frac{\partial^2 \bar{S}_n}{\partial\mat{x}'\partial\mat{x}''} \\
				\frac{\partial^2 \bar{S}_n}{\partial\mat{x}''\partial\mat{x}'} &
				\frac{\partial^2 \bar{S}_n}{\partial\mat{x}''\partial\mat{x}''}
			\end{pmatrix}
			\right.
			-
			\left.
			\varepsilon_{n n'}
			\begin{pmatrix}
				\frac{\partial^2 \bar{S}_n}{\partial\mat{x}'\partial \tau_n} \\
				\frac{\partial^2 \bar{S}_n}{\partial\mat{x}''\partial \tau_n}
			\end{pmatrix}
			\left(\pder[2]{\bar{S}_0}{\tau_0} + \pder[2]{\bar{S}_1}{\tau_1}\right)^{-1}
			\begin{pmatrix}
				\frac{\partial^2 \bar{S}_{n'}}{\partial \tau_{n'} \partial \mat{x}'} &
				\frac{\partial^2 \bar{S}_{n'}}{\partial \tau_{n'} \partial \mat{x}''}
			\end{pmatrix}
			\right|
\\
		&= \left| \begin{matrix} \frac{\partial^2 \bar{S}}{\partial\mat{x}'\partial\mat{x}'} &
										 \frac{\partial^2 \bar{S}}{\partial\mat{x}'\partial\mat{x}''} &
										 \frac{\partial^2 \bar{S}}{\partial\mat{x}' \partial\tau} \\
										 \frac{\partial^2 \bar{S}}{\partial\mat{x}''\partial\mat{x}'} &
										 \frac{\partial^2 \bar{S}}{\partial\mat{x}''\partial\mat{x}''} &
										 \frac{\partial^2 \bar{S}}{\partial\mat{x}'' \partial\tau} \\
										 \frac{\partial^2 \bar{S}}{\partial\tau \partial\mat{x}'} &
										 \frac{\partial^2 \bar{S}}{\partial\tau \partial\mat{x}''} &
										 \frac{\partial^2 \bar{S}}{\partial\tau^2} \\
						 \end{matrix} \right|,
\end{align}
and
$\varepsilon_{n n'}=2\delta_{n n'}-1$ is a permutation symbol taking values $\pm1$.
\end{widetext}

We are now able to re-express \eqn{kWpartial}
as
\begin{align}
	k_\text{SC} Z_0
	&=
			\sqrt{2\pi\hbar} \, \frac{\Delta^2}{\hbar^2}
			\sqrt\frac{\bar{C}_0 \bar{C}_1}{-\Sigma}
			\, \eu{-\bar{S}/\hbar},
	\label{kSpartial}
\end{align}
or
following a similar transformation to that of \eqn{d2WdE2}, with $\bar{W}$ and $E$ replaced by $\bar{S}$ and $\tau$, as
\begin{align}
	k_\text{SC} Z_0
	&=
			\sqrt{2\pi\hbar} \, \frac{\Delta^2}{\hbar^2}
			\sqrt{\frac{\bar{C}_0\bar{C}_1}{\bar{C}}}
			\left(-\frac{\rmd^2 \bar{S}}{\rmd\tau^2}\right)^{-\half} 
			\,\eu{-\bar{S}/\hbar}
	\label{kSfull}
\\
	\bar{C} &= \left| \begin{matrix} \frac{\partial^2 \bar{S}}{\partial\mat{x}'\partial\mat{x}'} &
										 \frac{\partial^2 \bar{S}}{\partial\mat{x}'\partial\mat{x}''} \\
										 \frac{\partial^2 \bar{S}}{\partial\mat{x}''\partial\mat{x}'} &
										 \frac{\partial^2 \bar{S}}{\partial\mat{x}''\partial\mat{x}''}
						 \end{matrix} \right|. %
\end{align}
This is very similar to the golden-rule instanton formulation of Cao and Voth \cite{Cao1997nonadiabatic}
which was derived from an extension to \ImF\ theory.
The only difference is that an extra approximation to the prefactor was made,
valid only for the spin-boson model (see \secref{spinboson}),
that the determinants cancel with the partition function
to give
\begin{align} \label{Voth}
	k_\text{C\&V}
	= \sqrt{2\pi\hbar} \, \frac{\Delta^2}{\hbar^2}
			 \left(-\frac{\rmd^2 \bar{S}}{\rmd\tau^2}\right)^{-\half} 
			\eu{-\bar{S}/\hbar},
\end{align}
where again $\tau$ is given as the value which solves $\pder{\bar{S}}{\tau}=0$.
However, had this approximation not been taken,
the two approaches would have given exactly equivalent results.
It is interesting to see that the current approach gives the same result as an \ImF\ formulation which
was applied to nonadiabatic problems without rigorous derivation.
The work presented in this paper, which gives an equivalent result, thus provides an explanation of why the
golden-rule instanton formulation of Cao and Voth recovered the semiclassical result
for the spin-boson model, \cite{Cao1997nonadiabatic}
and would also apply to more general systems if the extra approximation had not been made.
In fact, this shows that the \ImF\ approach 
works remarkably well in the golden-rule limit.
However, the new derivation based on semiclassical Green's functions is simpler
as it does not involve analytic continuation of divergent integrals
and starts from a rigorous energy-space picture of the reaction.

Generalizations of this nonadiabatic instanton approach
have been proposed
which interpolate the electron-transfer rate between the weak- and strong-coupling limits.
\cite{Cao1995nonadiabatic,Schwieters1998diabatic}
However, they were also based on extensions of the \ImF\ approach,
which does not appear to lead to a unique formulation.
In one case, the instanton was projected onto pure diabatic states, which causes errors in the adiabatic limit, \cite{Cao1995nonadiabatic}
whereas in the other, all electronic configurations were allowed
giving a mean-field approach which would fail to describe the high-temperature golden-rule rate. \cite{Schwieters1998diabatic}
when the instanton shrinks.

\section{Analytic solution in special cases}
\label{sec:analytic}

It is possible to find a closed-form expression of the golden-rule instanton rate %
in a few special cases.
In particular, we treat a system with linear potentials, the spin-boson model and the classical limit for a general condensed-phase electron transfer.

\subsection{One-dimensional linear potentials}
\label{sec:linear}

The simplest description of a nonadiabatic curve crossing problem
is that of two linear potentials,
\begin{align}
	V_n(x) &= V^\ddag + \kappa_n x
\end{align}
with $\kappa_0 > 0 > \kappa_1$ and a constant coupling, $\Delta$.
This is similar to the famous Landau-Zener model \cite{Zener1932LZ}
but without the constraint that the particle travels at a constant speed along the $x$ coordinate.
In order to define a rate, we assume that the potential only has this linear form near the crossing point
and flattens out at extreme values of $x$
without affecting the transmission probability for relevant values of energy.
The reactant partition function $Z_0$ could then be defined
using the translational invariance of incoming scattering states.

Using the wave functions given in \secref{greens},
and noting that the integrals over the Airy functions can be performed analytically
using \Ref{Aspnes1966Airy},
gives the exact transmission probability
\begin{multline}
	P(E) = 4\pi^2\Delta^2 \left(\frac{-4m^2}{\hbar^4\kappa_0(\kappa_0-\kappa_1)^2\kappa_1}\right)^{1/3}
	\\\times
	\Ai^2\left[-\left(\frac{2m(\kappa_0-\kappa_1)^2}{\hbar^2\kappa_0^2\kappa_1^2}\right)^{1/3}E\right].
\end{multline}
The golden-rule instanton approach, \eqn{PSC}, gives for $E<V^\ddag$,
\begin{multline}
	P_\text{SC}(E) = \frac{\pi\Delta^2}{\hbar|\kappa_0-\kappa_1|} \sqrt\frac{2m}{V^\ddag-E}
	\\\times \exp\frac{4\sqrt{2m}(V^\ddag-E)^{3/2}|\kappa_0-\kappa_1|}{3\hbar\kappa_0\kappa_1}.
\end{multline}
Using the asymptotic limit, \eqn{approxAi},
shows that this tends to the exact result as $E\rightarrow-\infty$.
It diverges however at $E=V^\ddag$ and is not applicable for higher energies.

\begin{figure}
	\includegraphics{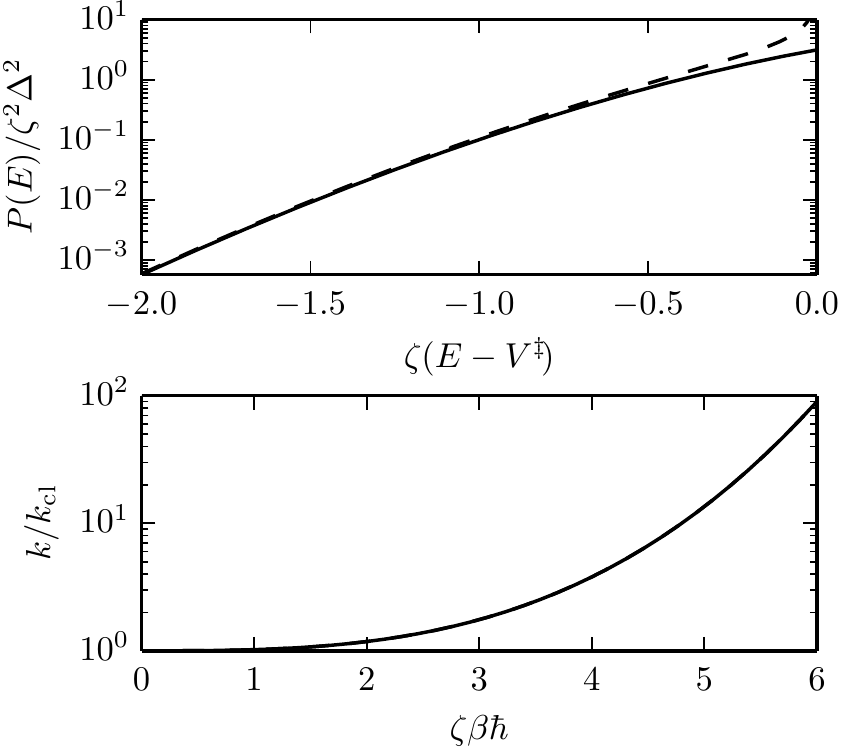}
	\caption{Transmission coefficient and thermal rate constants for the one-dimensional linear model.
	Solid lines are exact results and dashed from instanton approximation.
	Energies and temperatures are weighted by
	$\zeta=(2m/\kappa^2\hbar^2)^{1/3}$
	to give dimensionless units.
	}
	\label{fig:rate}
\end{figure}

For the thermal rate, defined by \eqn{kSpartial}, we obtain
\begin{align}
k_\mathrm{SC} = k_\mathrm{cl} \exp\frac{\beta^3\hbar^2\kappa_0^2\kappa_1^2}{24m(\kappa_0-\kappa_1)^2},
\end{align}
where the classical rate is \cite{nonoscillatory}
\begin{align}
	k_\mathrm{cl}Z_0 &= \sqrt\frac{2\pi m}{\beta\hbar^2} \frac{\Delta^2}{\hbar|\kappa_0-\kappa_1|} \, \eu{-\beta V^\ddag}.
\end{align}
We plot these results in \fig{rate} for the symmetrical case $\kappa\equiv\kappa_0=-\kappa_1$.
As expected the semiclassical results deviate from the exact transmission probability
when the energy nears the diabatic crossing $V^\ddag$.
Interestingly however, there appears to be a cancellation of errors
as the thermal rate gives the true value perfectly.
This is confirmed formally
using the integral identity
$\int_{-\infty}^\infty \Ai^2(-aE) \, \eu{-\beta E} \, \rmd E = \eu{\beta^3/12a^3} / \sqrt{4\pi a\beta}$,
for $a,\beta>0$.
The same result is of course obtained using the golden-rule flux correlation function \cite{Miller1983rate,nonoscillatory}
and the exact path-integral propagator. \cite{Feynman,Brown1994linear}

The figure shows that tunnelling changes the rate by orders of magnitude at low temperature
but that the classical result is correct in the high-temperature limit.
This simple example could be used to give a rough estimate of the importance of nuclear tunnelling
in more complex systems if a one-dimensional reaction coordinate is known.
However this is not generally the case and the full-dimensional instanton pathway will normally need to be found
for accurate predictions.

\subsection{Spin-boson model}
\label{sec:spinboson}

The standard model for electron-transfer reactions in the condensed phase
is the spin-boson model. \cite{Garg1985spinboson,*Leggett1987spinboson,Weiss}
The potentials of reactants and products are given by sets of shifted harmonic oscillators,
\begin{subequations}
\begin{align}
	V_0(\mat{x}) &= \sum_{j=1}^f \half m \omega_j^2 (x_j + \xi_j)^2  \\
	V_1(\mat{x}) &= \sum_{j=1}^f \half m \omega_j^2 (x_j - \xi_j)^2 - \epsilon,
\end{align}
\end{subequations}
and the electronic coupling, $\Delta$, is taken to be a constant.
The frequencies and couplings of the modes are determined according to a
given spectral density, which we write the discretized form,
$J(\omega) = \frac{\pi}{2} \sum_j \frac{c_j^2}{m\omega_j} \delta(\omega-\omega_j)$,
where $c_j=m\omega_j^2\xi_j$.

The exact quantum golden-rule rate for this system can be calculated numerically using
\cite{Weiss}
\begin{align}
	k_\text{QM} &= \frac{\Delta^2}{\hbar^2} \int_{-\infty-\iu\tau}^{\infty-\iu\tau} \eu{-\phi(t)/\hbar} \, \rmd t \\
	\phi(t) &= - \iu \epsilon t + \frac{4}{\pi} \int \frac{J(\omega)}{\omega^2} \left[ \frac{1-\cos{\omega t}}{\tanh\half\beta\hbar\omega} + \iu\sin{\omega t} \right] \rmd\omega,
\end{align}
where $\tau$ can be chosen to simplify the numerical integral over $t$ as much as possible
but has no effect on the result.

A common approximation to the quantum golden-rule rate takes
a stationary-phase integral about the point $t=-\iu\tau$ such that $\phi'(-\iu\tau)=0$.
This gives \cite{Weiss}
\begin{align}
	k_\text{SP} &= \frac{\Delta^2}{\hbar^2} \sqrt\frac{2\pi\hbar}{\phi''(-\iu\tau)} \, \eu{-\phi(-\iu\tau)/\hbar}.
	\label{SP}
\end{align}
Similar formulations exist for generalizations of the spin-boson model
to include anharmonicities, \cite{Tang1994anharmonic}
although practical expressions
are limited to system-bath problems of a particular form
where the bath can be represented by an effective harmonic form. \cite{Makri1999linearresponse}

We now proceed to compute the golden-rule instanton rate for this system.
The Euclidean actions for classical trajectories in this harmonic system are
\cite{Feynman,Kleinert}
\begin{widetext}
\begin{subequations}
\begin{align}
	\bar{S}_0(\mat{x}',\mat{x}'',\tau_0)
		&= \sum_{j=1}^f \frac{m \omega_j}{2\sinh{\omega_j\tau_0}}
		\left[ \big((x_j' + \xi_j)^2 + (x_j'' + \xi_j)^2\big)\cosh{\omega_j\tau_0} 
		- 2(x_j' + \xi_j)(x_j'' + \xi_j) \right]
\\
	\bar{S}_1(\mat{x}',\mat{x}'',\tau_1)
		&= \sum_{j=1}^f \frac{m \omega_j}{2\sinh{\omega_j\tau_1}}
		\left[ \big((x_j' - \xi_j)^2 + (x_j'' - \xi_j)^2\big)\cosh{\omega_j\tau_1} 
		- 2(x_j' - \xi_j)(x_j'' - \xi_j) \right]
		- \epsilon\tau_1.
\end{align}
\end{subequations}
\end{widetext}

Solving $\pder{\bar{S}}{\mat{x}'}=\pder{\bar{S}}{\mat{x}''}=0$,
we find that
\begin{align}
	x_j^\ddag = -\xi_j \frac{\sinh{\half\omega_j(\beta\hbar-2\tau)}}{\sinh{u_j}},
\end{align}
where $u_j = \beta\hbar\omega_j/2$.
Therefore the periodic orbit which hops at this point has action
\begin{align}
	\bar{S}(\tau) = - \epsilon\tau
		+ \sum_{j=1}^f 2 m \omega_j \xi_j^2 \left[\frac{1-\cosh{\omega_j\tau}}{\tanh{u_j}} + \sinh{\omega_j\tau}\right],
\end{align}
which is simply $\phi(-\iu\tau)$ for an $f$-dimensional system. %
We must in general 
choose $\tau$ numerically to solve $\pder{\bar{S}(\tau)}{\tau}=0$
to ensure that the energy of the two Green's functions are equal.
In the unbiased case where $\epsilon=0$, symmetry considerations give $\tau=\half\beta\hbar$.
However, we continue with the proof in the general case.

The second derivatives are
\begin{subequations}
\begin{align}
	\frac{\partial^2\bar{S}_n}{\partial x_j' \partial x_k'}
		&= \delta_{jk} \frac{m \omega_j}{\tanh{\omega_j\tau_n}} \\
	\frac{\partial^2\bar{S}_n}{\partial x_j' \partial x_k''}
		&= - \delta_{jk} \frac{m \omega_j}{\sinh{\omega_j\tau_n}}
\end{align}
\end{subequations}
and their equivalents with all single and double primes exchanged.
This gives
\begin{align}
	\bar{C}_n
	&= \prod_j \frac{m\omega_j}{\sinh{\omega_j\tau_n}}
\\
	\bar{C}
	&= \prod_j m^2\omega_j^2 \left(2 + \frac{\tanh{\half\omega_j\tau_0}}{\tanh{\half\omega_j\tau_1}} + \frac{\tanh{\half\omega_j\tau_1}}{\tanh{\half\omega_j\tau_0}}\right)
\\
	&= \frac{\bar{C}_0 \bar{C}_1}{Z_0^2},
\end{align}
where we have used \eqn{detequal} as well as a number of trigonometric relations.
For this case of the spin-boson model, the determinants in \eqn{kSfull} cancel with the partition function, \eqn{Z0},
and we may proceed with the formula of Cao and Voth, \eqn{Voth}, without approximation.
It is then seen that our instanton approach gives the same result as the stationary-phase approximation, \eqn{SP}, for the spin-boson model.
Note that for this harmonic system,
the fluctuations about the instanton pathway are treated exactly
and so the same result would also be obtained by the quantum instanton method of Wolynes. \cite{Wolynes1987nonadiabatic}

It was shown for example in \Ref{Bader1990golden} that this stationary-phase approximation introduces an error of about 20\%
for a model of aqueous ferrous-ferric electron transfer. 
Such an error is quite acceptable as often only the order of magnitude of the rate is required.
The stationary-phase approximation, and thus the golden-rule instanton, is not equivalent to the ``semiclassical'' method
described in \Ref{Siders1981quantum}.
In the latter case, an extra approximation was made that $\op{H}_0$ and $\op{H}_1$ commute
and thus gives inferior results at low temperatures. \cite{Bader1990golden}

The stationary-phase approximation to the golden-rule rate in the spin-boson model
becomes exact in the classical (high-temperature) limit and recovers Marcus theory. \cite{Marcus1985review,Weiss}
It is interesting to note that adiabatic instantons
tend to an asymptotic low-temperature limit
but break down at a certain cross-over temperature
and a different form is needed in the high temperature regime.
\cite{*[{An extension for the cross-over regime is outlined in }] [{}] Zhang2014interpolation}
This does not occur in the instanton approximation for the golden-rule rate,
not just for the spin-boson model, but for any system
as we shall show in the next section,
which deals with
the high-temperature limit more generally. %

\subsection{Classical limit}
\label{sec:classical}

We now consider a limit which has not been previously well studied,
which is the classical limit for the golden-rule rate of a generic anharmonic system.
We recently derived a general formula for the classical golden-rule rate \cite{nonoscillatory}
and we will show here that the semiclassical instanton method tends to a steepest-descent version of it.

To achieve the classical, high-temperature limit, we let $\beta\rightarrow0$
which forces the periodic orbit to become an infinitesimally short line.
As before we require that the hop occurs at $\mat{x}'=\mat{x}''=\mat{x}^\ddag$ where the potentials are equal,
and the trajectories must bounce along the direction of the gradient and be continuous at the hop.
It is therefore possible to conclude that
at the hopping point, $\mat{x}^\ddag$,
the directions of the two gradients are exactly opposite
and it is thus the minimum on the crossing seam, $V_0(\mat{x})=V_1(\mat{x})$.
We are therefore allowed to describe the potentials
using the following series expansion about this point:
\begin{align}
	V_n(\mat{x}) &\approx V^\ddag + \mat{g}_n^\T (\mat{x} - \mat{x}^\ddag) + \half m (\mat{x}-\mat{x}^\ddag)^\T \mat{\Omega}_n (\mat{x}-\mat{x}^\ddag).
\end{align}
We choose an orthogonal coordinate system
such that the degree of freedom $j=1$ is normal to the crossing seam at the hopping point, $\mat{x}^\ddag$.
All other modes are at their minimum positions here such that $\mat{g}_n=(\kappa_n,0,\dots,0)$.
Note that this \textit{local} expansion is by no means equivalent to the \textit{global} harmonic approximation employed by the spin-boson model and Marcus theory.
Here for instance, the frequencies at the crossing point are allowed to differ on each electronic surface as well as in the reactant well.
As the instanton method is derived by performing steepest-descent integrations,
including higher order terms in the expansion would not change the following results.
We thus expect to obtain the correct activation energy even when the reorganization energy of the reactants and products are unequal.

In the short-time limit,
\begin{align}
	\bar{S}_n(\mat{x}',\mat{x}'',\tau_n) = \frac{m}{2\tau_n}|\mat{x}'-\mat{x}''|^2 + V_n\left(\half(\mat{x}'+\mat{x}'')\right)\tau_n,
\end{align}
and we find that in order for the total action to be stationary with respect to $\mat{x}^\ddag$,
\begin{align}
	\tau_0 &= \beta\hbar \frac{\kappa_1}{\kappa_1-\kappa_0}, &
	\tau_1 &= \beta\hbar \frac{\kappa_0}{\kappa_0-\kappa_1}.
\end{align}

The fluctuation terms tend to
\begin{align}
	\bar{C}_n &= \left(\frac{m}{\tau_n}\right)^f,
\end{align}
and
due to symmetry between $\mat{x}'$ and $\mat{x}''$ we can 
simplify the determinant by
rotating the axes to $(\mat{x}''+\mat{x}')/\sqrt{2}$ and $(\mat{x}''-\mat{x}')/\sqrt{2}$.
This is equivalent to using \eqn{detsymmetry},
and gives
\begin{align}
	\Sigma
	&= \left| \begin{matrix} 2m \left(\frac{1}{\tau_0} + \frac{1}{\tau_1}\right) \mathbb{1} &
								 \mathbb{0} &
								 \mat{0} \\
								 \mathbb{0} &
							    \half \beta \hbar m \tilde{\mat{\Omega}} &
								 \frac{1}{\sqrt{2}} (\mat{g}_1 - \mat{g}_0) \\
								 \mat{0}^\T &
								 \frac{1}{\sqrt{2}} (\mat{g}_1 - \mat{g}_0)^\T &
								 0
				 \end{matrix} \right|
\\
	&= \left[\frac{2m}{\beta\hbar} \frac{(\kappa_0-\kappa_1)^2}{-\kappa_0\kappa_1}\right]^f
		\left| \half \beta \hbar m \tilde{\mat{\Omega}} \right|
		\nonumber\\&\qquad\times
		\left( - \frac{1}{\beta \hbar m} (\mat{g}_1 - \mat{g}_0)^\T \tilde{\mat{\Omega}}^{-1} (\mat{g}_1 - \mat{g}_0) \right)
\\
	&= - \left[m^2 \frac{(\kappa_0-\kappa_1)^2}{-\kappa_0\kappa_1}\right]^f
		\frac{(\kappa_1 - \kappa_0)^2}{\beta \hbar m}
		|\tilde{\mat{\Omega}}|_{11},
\end{align}
where $\tilde{\mat{\Omega}} = \frac{\kappa_0\mat{\Omega}_1 - \kappa_1\mat{\Omega}_0}{\kappa_0 - \kappa_1}$
and the minor $|\tilde{\mat{\Omega}}|_{11}$ is formed by removing the first row and column of the matrix and taking the determinant;
it is defined to be equal to 1 in the case that $f=1$.
Therefore the golden-rule instanton method approaches the following form in the classical limit:
\begin{align}
	\label{clhTST}
	k_\mathrm{cl,hTST}
		&= \sqrt\frac{2\pi m}{\beta\hbar^2} \frac{\Delta^2}{\hbar|\kappa_0 - \kappa_1|}
		\frac{Z^\ddag}{Z_0}
		\, \eu{-\beta V^\ddag}
\\
		Z^\ddag &= |\beta^2\hbar^2\tilde{\mat{\Omega}}|_{11}^{-1/2}.
\end{align}
For a one-dimensional system, $Z^\ddag=1$ and the formula is
obviously equal to the classical golden-rule TST rate. \cite{UlstrupBook,nonoscillatory,Tang1994anharmonic}
\Eqn{clhTST} is proportional to this one-dimensional formula evaluated along a particular reaction coordinate,
which is defined as being normal to the crossing seam.
This rate is simply multiplied by classical fluctuations (vibrational partition functions in the harmonic approximation) along perpendicular coordinates,
as expressed by the minor of the matrix.
We name this harmonic classical golden-rule TST and note that it recovers Marcus theory when applied to the spin-boson model.
It therefore makes the same assumption that the reaction coordinate can be separated
as is used in Born-Oppenheimer classical TST, \cite{Miller1974QTST}
and is therefore the golden-rule equivalent of this.
A simple extension would give the equivalent to Eyring TST \cite{Eyring1938rate}
by replacing the classical partition functions in $Z^\ddag$ by their quantum versions.
However such a method is not as accurate as the instanton approach as it still treats motion along the reaction coordinate with classical dynamics.

It is a major success of the golden-rule instanton theory presented here
that we recover the classical limit for high temperatures.
This is however not unexpected
as we have started from exact golden-rule expression and performed a semiclassical approximation
thus preserving the term of lowest order in $\hbar$, which of course gives the correct steepest-descent classical result.
Semiclassical instanton methods
developed to study single-surface reactions \cite{Miller1975semiclassical,rpinst} %
break down at a particular crossover temperature
where the instanton collapses to a singularity
and require a different formula for the high-temperature limit.
This occurs because the potential at the barrier tends to a parabola with finite curvature
which cannot support short periodic orbits above a certain cross-over temperature. \cite{Benderskii,rpinst}
However, the golden-rule instanton never collapses completely because the potentials become approximately linear near the activation energy
forming a cusp,
and at least a small amount of tunnelling occurs at all temperatures.

We also note that the extra approximation taken by Cao and Voth, \cite{Cao1997nonadiabatic}
gives a method, \eqn{Voth},
which assumes that the normal modes at the minimum of the reactant potential
are equal to those at the transition state on each surface.
This therefore only gives the correct classical limit for the case of the spin-boson model.

\section{Conclusions}
\label{sec:conclusions}

In this paper we have derived a semiclassical instanton method for computing the rate of an electron-transfer reaction in the nonadiabatic limit.
Our derivation
starts from an exact expression for the golden-rule rate written in terms of Green's functions,
which are themselves approximated by a semiclassical limit
based on classical trajectories.
The remaining integrals are evaluated within the steepest-descent method.
This procedure defines
two imaginary-time trajectories which contribute to the rate,
one on each of the reactant and product potential-energy surfaces,
both of which have the same energy and must encounter a bounce.
The hop between the surfaces occurs at a point where the potentials are equal
and the trajectories join together smoothly to make a periodic instanton orbit with imaginary time $\beta\hbar$.

In this way, we have derived four equivalent formulae
to define the golden-rule instanton rate:
\eqs{kWfull}, \eqref{kWpartial}, \eqref{kSpartial} and \eqref{kSfull},
which use either a Hamilton-Jacobi or Lagrangian formulation
and either full or partial derivatives with respect to energy or imaginary time.
In Paper II, which follows this article,
we shall show how these formulae can be evaluated numerically for a complex multidimensional system.

In our opinion this derivation
makes clear the assumptions and approximations being made
and
provides physical insight into the electron-transfer process,
showing that the dominant contribution comes from an instanton of constant energy which hops on the diabatic crossing seam.
Previous golden-rule instanton formulae were based on the \ImF\ approach \cite{Cao1997nonadiabatic}
and it is interesting to see how closely the methods relate to each other,
considering that their derivations are apparently so different.

The golden-rule instanton approach
gives the exact thermal rate for a system with linear potentials
and, 
for the case of the spin-boson model,
reproduces the well-known stationary-phase evaluation of the quantum golden-rule formula, \eqn{SP}.
Note that although the steepest-descent approach uses only harmonic fluctuations about the dominant instanton pathway,
the important integral along the instanton path is evaluated exactly.
The remaining approximation is equivalent only to a \textit{local} rather than a \textit{global} harmonic approximation.
In other words the harmonic frequencies are allowed to vary along the instanton pathway
and the method should also give good results for anharmonic systems,
including problems where the reorganization energies of products and reactants are different.

In the high-temperature classical limit,
it reduces in general to a steepest-descent form of classical golden-rule TST \cite{nonoscillatory}
and thus to Marcus theory for the spin-boson model.
It can therefore be seen as an extension of classical golden-rule TST to the quantum regime.
However, unlike the adiabatic instantons
there is some tunnelling at all temperatures
and no crossover from the deep to shallow-tunnelling regimes. %
This shows that nuclear quantum effects are always apparent when computing the golden-rule rate
and therefore that such methods will be necessary for the accurate simulation of electron transfer.

The method is valid for any system with a simple potential-energy landscape,
where either
only one instanton exists
or there are multiple instantons which are well separated.
However, all instanton approaches will fail when the environment is fluxional,
as is the case for electron-transfer in solution.
Nonetheless, a semiclassical instanton analysis can often
be used to better understand the approximations involved in related methods.
An example is the link between adiabatic instantons and RPTST, \cite{rpinst}
which has been shown to be exact in the absence of recrossing. \cite{Hele2013QTST,Althorpe2013QTST,*Hele2013unique}
A related semiclassical study of electron-transfer pathways \cite{Shushkov2013instanton}
has also helped improve attempts to describe nonadiabatic dynamics. \cite{Menzeleev2011ET,Menzeleev2014kinetic}
In the following paper, \cite{GoldenRPI}
we show how Wolynes' quantum instanton method \cite{Wolynes1987nonadiabatic}
is related to our approach.

When computing the semiclassical Green's functions,
we only considered imaginary-time trajectories
and have therefore ignored real-time effects.
For this reason the instanton approach should be considered to be a (quantum) transition-state theory.
In the study of adiabatic reaction rates, it has been shown that
ring-polymer molecular dynamics (RPMD) \cite{Habershon2013RPMDreview,Hele2015Matsubara,*Hele2015RPMD}
generalizes RPTST and hence the adiabatic instanton. \cite{rpinst}
Taking inspiration from this,
it may be possible to combine such TST instanton methods with nonadiabatic RPMD \cite{mapping,Ananth2013MVRPMD}
to compute recrossing effects.

The semiclassical Green's function formalism is not limited to treating golden-rule rate problems.
In forthcoming publications
we will show how similar approaches
can be applied to the Marcus inverted regime
and to compute the rate constant 
in the adiabatic, large $\Delta$ limit.

\section{Acknowledgement}
JOR gratefully acknowledges a Research Fellowship from the Alexander von Humboldt Foundation.

\appendix

\section{Properties of determinants}
\label{determinant}
We shall make frequent use of the following expansions of the determinant of a block matrix:
\begin{align}
	\label{detblock}
	\left|\begin{array}{cc} \mat{A} & \mat{B} \\ \mat{C} & \mat{D} \end{array} \right|
	= |\mat{D}| \left| \mat{A} - \mat{C} \mat{D}^{-1} \mat{B} \right|
	= |\mat{A}| \left| \mat{D} - \mat{B} \mat{A}^{-1} \mat{C} \right|,
\end{align}
which are valid as long as the inverses of $\mat{D}$ or $\mat{A}$ exist.

When certain blocks in the determinant are equal,
and $\mat{A}$, $\mat{B}$ and $\mathbb{0}$ are square matrices of the same size,
$\mat{C}$ and $\mat{0}$ are vectors of the same length and $D$ is a scalar,
the following simplifications are found by rotating the basis set:
\begin{align}
	\label{detsymmetry}
	\left|\begin{array}{ccc} \mat{A} & \mat{B} & \mat{C} \\ \mat{B} & \mat{A} & \mat{C} \\ \mat{C}^\T & \mat{C}^\T & D \end{array} \right|
	&=
	\left|\begin{array}{ccc} \mat{A}-\mat{B} & \mathbb{0} & \mat{0} \\ \mathbb{0} & \mat{A}+\mat{B} & \sqrt{2}\mat{C} \\ \mat{0}^\T & \sqrt{2}\mat{C}^\T & D \end{array} \right|,
\end{align}
and also
\begin{align} 
	\label{detequal}
	\left|\begin{array}{cc} \mat{A} & \mat{B} \\ \mat{B} & \mat{A} \end{array} \right|
	&= \left|\mat{A} - \mat{B}\right| \left| \mat{A} + \mat{B} \right|.
\end{align}

\bibliography{references} %

\end{document}